\documentclass[11pt]{article}

\usepackage[T1]{fontenc}
\usepackage[latin1]{inputenc}
\usepackage{fullpage}
\usepackage{url}
\usepackage{amsmath}
\usepackage{amssymb}
\usepackage{amsfonts}
\usepackage{amsthm}
\usepackage{enumerate}
\usepackage{epsfig}
\usepackage{color}
\usepackage{xypic}
\xyoption{all}

\newcommand{\mc}{\mathcal}
\newcommand{\cqfd}{\hfill\ensuremath{\Box}}
\newcommand{\bec}{\begin{center}}
\newcommand{\eec}{\end{center}}
\newcommand{\btab}{\begin{tabular}}
\newcommand{\etab}{\end{tabular}}
\newcommand{\ie}{i.e.\ }

\newcommand{\begit}{\begin{enumerate}[$\bullet$]}
\newcommand{\eeit}{\end{enumerate}}

\newcommand{\dd}{\textrm{d}}

\newcommand{\proba}{\mathcal{P}}

\newcommand{\maximum}{\textrm{max}}

\newcommand{\MaxPos}{\textrm{MaxPos}}
\newcommand{\Price}{\textrm{Price}}

\newcommand{\linearise}{\mathfrak{L}}

\newcommand{\piSol}{\pi^*}
\newcommand{\pproba}{P}
\newcommand{\pathInt}{\mathcal{D}}
\newcommand{\pathIntPred}{\mathcal{L}}
\newcommand{\probaCross}{\mathcal{P}}
\newcommand{\qstar}{{q^*}}

\newcommand{\indicFun}{\mathbf{1}}
\newcommand{\pRef}{\mathfrak{p}}
\newcommand{\gspecial}{\overline{g}}
\newcommand{\diffPos}{\Delta\pi}

\title{Optimal Trading with Linear Costs}
\author{Joachim de Lataillade, Cyril Deremble,\\
  Marc Potters \& Jean-Philippe Bouchaud\\
  Capital Fund Management, 6 Boulevard Haussmann, 75009 Paris, France}
\date{}

\begin{document}

\newtheorem{Proposition}{Proposition}

\maketitle

\begin{abstract}
  We consider the problem of the optimal trading strategy in the
  presence of linear costs, and with a strict cap on the allowed
  position in the market. Using Bellman's backward recursion method,
  we show that the optimal strategy is to switch between the maximum
  allowed long position and the maximum allowed short position,
  whenever the predictor exceeds a threshold value, for which we
  establish an exact equation. This equation can be solved explicitely
  in the case of a discrete Ornstein-Uhlenbeck predictor. We discuss
  in detail the dependence of this threshold value on the transaction
  costs. Finally, we establish a strong connection between our problem
  and the case of a quadratic risk penalty, where our threshold
  becomes the size of the optimal non-trading band.
  % which leads us to recover the threshold increases as the one-third
  % power of these costs where we recover in a certain regime that
  % this threshold increases as the one-third power of these costs, a
  % result previously obtained for a quadratic risk penalty.  We show
  % that our result is in fact universal, independent of the specific
  % risk constraint.
 \end{abstract}

%\medskip

\section{Introduction}

Contrarily to the efficient market dogma, prices have some degree of
predictability, at least on short time scales.  Statistical arbitrage
strategies aim at exploiting this small predictability. However, costs
make it difficult, although not impossible, to eke out a profit from
these inefficiencies. Indeed, every trade is costly; the structure of
these costs is actually quite complex. Some of them are related to
various fees (market fees, brokerage fees, etc.)  and are usually a
small fraction of the traded quantity (typically $10^{-5}$ to
$10^{-4}$ on liquid markets).  We call these costs ``linear'', because
they are simply proportionnal to the traded amount. Another source of
linear costs is the bid-ask spread, which represents a few ``basis
points'' (bp, i.e. $10^{-4}$ of the value of the contract). Much more
subtle are impact induced costs, which come from the fact that a large
order must be split in a sequence of small trades that are executed
incrementally.  But since each executed trade, on average, impacts the
price in the direction of the trade, the average execution price is
higher (in the case of a buy) than the decision price, leading to what
is called ``execution shortfall''.  This cost is clearly non-linear,
since the price impact itself increases with the size $Q$ of the
trade. Empirical data suggests that the impact induced cost is on the
order of $\sigma Q^{3/2}/ \sqrt{V}$, where $\sigma$ is the daily
volatility and $V$ the daily turnover (see e.g. \cite{toth}). This shows that for an order
corresponding to $1 \%$ of the daily turnover, and for an asset with
$2 \%$ daily volatility, the impact cost is on the order of $20$
bp. But for much smaller orders, or for assets where the bid-ask
spread is large, linear costs can be dominant.

The problem we want to address and solve in this work is to determine
the optimal trading strategy when one has discovered a predictive
statistical signal, in the presence of linear trading costs and a
constraint on the maximum size of the position (both long and
short). While the case of a quadratic risk control has been considered
in the literature for linear costs~\cite{davisNorman, shreveSoner,
  cmeCosts}, quadratic costs (i.e. costs growing like $Q^2$)
\cite{strategiesImpact, almgrenLin} or various impact-dependent
cost~\cite{almgrenNonLin}, we are not aware of any published solution
in the case where the risk constraint is a cap on the position. This
problem can be of significant interest in practice, because the risk
of a trading system is sometimes handled completely outside of the
system, through such a cap on the position, in order to reduce
operational risk.

Our final solution for the optimal position $\pi$ as a function of the
predictor $p$ is as follows: the size of the position $|\pi|$ should
always be at the maximum allowed position $M$, with a sign that
switches between $-1$ and $+1$ whenever the predictor exceeds a {\it
  threshold value} $q^*$ (and vice-versa when the predictor becomes
smaller than $-q^*$). We find an exact equation for the threshold
$q^*$, that we solve explicitely in the case of a (discrete-time)
Ornstein-Uhlenbeck predictor.  In some limits we find that the
threshold $q^*$ scales as the one-third power of the cost parameter, a
result already discussed in the literature but in the context of a
quadratic risk contraint.  We explain why our result is strongly
connected with this alternative problem. We also check the validity of
our results using numerical simulations.

% We explain why our result is in fact {\it universal}, and
% independent of the particular shape of the risk constraint. We also
% check the validity of our results using numerical simulations.

\section{Description of the problem}\label{description}

We consider an agent who trades a single asset, of current price
$\Price_t$. The position (signed number of shares/contracts) of the trader at
time $t$ is $\pi_t$. We assume that the agent has some signal $p_t$
that predicts the next price change $r_t = \Price_{t+1}-\Price_t$, and
is faced with the following constraints:
\begin{itemize}
\item His/her risk control system is simply a cap on the absolute size
  of his position~: $|\pi|\leq M$, with no other risk control. $M$
  will be called the ``\MaxPos'' of the agent.
\item He/she has to pay linear costs $\Gamma Q$ whenever he/she trades a
  quantity $Q=|\delta{\pi}|=|\pi_{t+1}-\pi_t|$
\end{itemize}

The agent wants to maximise his/her expected gains, by trading over a
long period $[0, T]$ (we will later consider the limit
$T\rightarrow\infty$).

We also assume the predictor to~:
\begin{enumerate}[(i)]
\item\label{linearPred} have a linear predictability:
  $E[r_t|p_t]=A\cdot p_t$ with some constant $A$
\item be positively auto-correlated: $\forall q$,
  $\pproba(p_{t+1}>q|p_t)$ increases continuously with $p_t$
\item be Markovian: $\forall\omega_{t+1}$,
  $\pproba(\omega_{t+1}|p_t,p_{t-1},\dots)=\pproba(\omega_{t+1}|p_t)$ where $\omega_{t+1}$ is any event at $t+1$
\item be symmetric: $\forall q,\ \pproba(p_{t+1}>q|p_t=p)\ =\
  \pproba(p_{t+1}<-q|p_t=-p)$
\item\label{unbound} be unbounded: $\forall q,\ \exists
  \epsilon_q>0\textrm{ s.t. }\pproba(p_{t+1}>q|p_t=0)\ >\epsilon_q$.
\end{enumerate}

\noindent\textbf{Remarks:}
\begin{itemize}
\item Without loss of generality, we can always set $A=1$, so that
  $p_t$ is in price units.
  % \item Condition~(\ref{unbound}) is not compulsory: in
  %   Section~\ref{pathIntegral} we will express the solution of the
  %   problem with and without that condition. This condition ensures
  %   that there is always a possibility to generate a positive gain
  %   with the predictor, whatever the costs.
\item Because of the positive auto-correlation of $p_t$, we can define
  an \textbf{integrated predictability} at $t=\infty$, depending on
  $p_t$~:
$$p_\infty(p_t)=E[\ \Price_{\infty}-\Price_t\ |\ p_t\ ] =\sum_{n=0}^\infty\ E[\
p_{t+n}\ |\ p_t\ ]$$ This quantity indicates how much one will gain in
the future if one keeps a fixed position $\pi_{t' \ge t}=\pi$~: the
expected gain is then $p_\infty(p_t)\ \pi$.
\item Finally, one can consider the case where
  Condition~(\ref{linearPred}) is not met, and
  note $$\linearise_t(p_t)=E[r_t|p_t]$$ the immediate
  predictability. Then, as we will see in
  Section~\ref{mostGeneralCase}, we can still express the solution of
  the system if we suppose that $\linearise_t$ is continuous, uneven,
  strictly increasing, and
  $\lim_{p_t\rightarrow\infty}\linearise_t(p_t)>\Gamma$. But the
  parameters of this solution will of course depend on these functions
  $\linearise_t$.

\end{itemize}

\section{The Optimal Strategy}\label{optimalStrat}

\subsection{A na\"\i ve solution}\label{naive}

At first sight, the solution to this problem seems straightforward: if
the expected future gain (given by the integrated predictability)
exceeds the trading cost per contract $\Gamma$, then one trades in the
direction of the signal (if not already at the \MaxPos), otherwise one
does not. This solution obviously generates a positive average gain,
but it has no reason to be the optimal solution. Indeed, because the
predictor is auto-correlated in time, it might be worthy (and in
general it will be) to wait for a larger value of the predictor, in
order to grab the opportunities that have the most chances to get
realised, and discard the others.  As we shall see, the mistake in
this na\"\i ve reasoning is not to compare the future gain with the
cost, but rather comes from a wrong definition of the future gain,
which does not include future trading decisions.

\subsection{The Bellman method}\label{solutionT}

To attack this problem, we will use Bellman's \textbf{optimal control
  theory}, or dynamic programming~\cite{bellman}, which consists in
solving the problem backwards: by assuming one follows the optimal
strategy for all future times $t'>t$, we can find the optimal solution
at time $t$. As is usual in dynamic programming, we have a
\textbf{control variable}\footnote{In the usual Bellman terminology
  the control variable is actually $\pi_t-\pi_{t-1}$, but here we want
  to insist on what we can control (the position) and what we cannot
  (the value of the predictor).} $\pi_t$, which needs to be optimised,
and a \textbf{state variable} $p_t$, which parametrises the solution.
The optimisation will be done through a \textbf{value function}
$V_t(\pi, p)$, which gives the maximal expected gains between time $t$
and $+\infty$, considering that the position at $t-1$ is $\pi$ and the
predictor's value at $t$ is $p$. The optimal solution of the system
will be denoted $(\piSol_t)_{t\in[0, T]}$.

\bigskip

Let us start with the simple case $t=T$, i.e. the optimal strategy at
the last time step. In this case the expected future return is really
$p_\infty(p) \pi_T$ where $p=p_T$, since no trading is allowed beyond
that time. Any trade $\delta\pi$ induces a cost $\Gamma|\delta\pi|$,
so:
\begin{itemize}
\item if $p_\infty(p)\geq\Gamma$ then $\piSol_T=M$,
  $V_T(\pi,p)=p_\infty(p)\cdot M - \Gamma(M - \pi)$.
\item if $p_\infty(p)\leq-\Gamma$ then $\piSol_T=-M$,
  $V_T(\pi,p)=-p_\infty(p)\cdot M - \Gamma(M + \pi)$.
\item if $|p_\infty(p)|<\Gamma$ then $\piSol_T=\pi$,
  $V_T(\pi,p)=p_\infty(p)\cdot\pi$.
\end{itemize}
Hence, we recover exactly the na\"\i ve solution in this case, but
this is only because there is no trading beyond $t=T$.

\bigskip

Now if we consider $t<T$, we have to maximise a quantity including
immediate gains, costs and future gains. This leads to the following
recurrence relation:
\begin{gather}\label{basicRecurrence}
  V_t(\pi,p)=\underset{|\pi'|\leq M}{\maximum}(\
  p\cdot\pi'-\Gamma|\pi'-\pi|+\int \pproba(p_{t+1}=p'\ |\ p_t=p)\
  V_{t+1}(\pi',p')\ \dd p'\ )
\end{gather}
and $\pi^*_t$ is the value of $\pi'$ which realises this maximimum when $\pi=\piSol_{t-1}$
and $p=p_t$.

\subsection{General solution}

In what follows, we will need the following notations:
\begin{gather*}
\pproba(p'|p)= \pproba(p_{t+1}=p'\ |\ p_t=p)\\
{\cal P}_{>q}(p)=\pproba(p_{t+1}>q\ |\ p_t=p)\\
{\cal P}_{<q}(p)=\pproba(p_{t+1}<q\ |\ p_t=p)
\end{gather*}
where the dependency on $t$ is kept implicit.

\medskip

\begin{Proposition}\label{generalSolution}
  There exist two functions $g(t,p)$ and $h(t,p)$ and a sequence
  $(q_t)_{t\in[0,T]}$ such that, for every $t\in[0, T]$, we have the
  following:
\begin{itemize}
\item $\piSol_t=\begin{cases}
\piSol_{t-1}&\textrm{if }|p_t|<q_t\\
M&\textrm{if }p_t\geq q_t\\
-M&\textrm{if }p_t\leq-q_t
\end{cases}$ \qquad (with $\pi_{-1}=0$)
\item $V_t(\pi, p)=\begin{cases}
g(t,p)\pi+h(t,p)&\textrm{if }|p_t|<q_t\\
(g(t,p)-\Gamma)\cdot M+\Gamma\pi+h(t,p)&\textrm{if }p_t\geq q_t\\
(-g(t,p)-\Gamma)\cdot M-\Gamma\pi+h(t,p)&\textrm{if }p_t\leq-q_t
\end{cases}$
%\end{gather*}
\item $g(t, p)$ is a continuous, strictly increasing function of $p$
  which satisfies, for $t<T$:
\begin{gather}\label{equation_g}
  g(t, p)=p+\Gamma\cdot[{\cal P}_{>q_{t+1}}(p)-{\cal
    P}_{<-q_{t+1}}(p)]+\int_{-q_{t+1}}^{q_{t+1}}\pproba(p'|p)g(t+1,
  p')\textrm{d}p'
\end{gather}
\item $q_t$ is such that $q_t\geq 0$ and
\begin{gather}
g(t, q_t)=\Gamma.
\end{gather}
\end{itemize}
\end{Proposition}

\bigskip

\emph{Proof.}  The proof is done by backwards recursion (\textit{ie.}
we assume that it is true for $t+1$ to prove that it is true for $t$).
One can easily check from section~(\ref{solutionT}) that the statement
is true for $t=T$, with in particular $g(T,p)=p_\infty(p)$.

Let us then suppose that it is true for $t+1$. We have the following:
\begin{align*}
  V_t(\pi,p)&=\underset{|\pi'|\leq M}{\maximum}\ (\
  p\cdot\pi'-\Gamma|\pi-\pi'|
  +\int \pproba(p'|p)\ V_{t+1}(\pi',p')\ \dd p'\ )\\
  &=\underset{|\pi'|\leq M}{\maximum}\ (\ p\cdot\pi'-\Gamma|\pi-\pi'|
  +\int_{-q_{t+1}}^{q_{t+1}}\pproba(p'|p)\cdot[g(t+1,p')\pi'+h(t+1,p')]\dd p'\\
  &\qquad\qquad\qquad+\int_{q_{t+1}}^{+\infty}\pproba(p'|p)\cdot[(g(t+1,p')-\Gamma)\cdot
  M
  +\Gamma\pi'+h(t+1,p')]\dd p'\\
  &\qquad\qquad\qquad+\int_{-\infty}^{-q_{t+1}}\pproba(p'|p)\cdot[(-g(t+1,p')-\Gamma)\cdot
  M-\Gamma\pi'+h(t+1,p')]\dd p'\ )
\end{align*}

So if we set
$$g(t, p)=p+\Gamma\cdot[{\cal P}_{>q_{t+1}}(p)-{\cal P}_{<-q_{t+1}}(p)]
+\int_{-q_{t+1}}^{q_{t+1}}\pproba(p'|p)g(t+1, p')\textrm{d}p'$$
and $$h(t,p)=\mc{H}\ +\
M\int_{q_{t+1}}^{+\infty}\pproba(p'|p)\cdot[g(t+1,p')-\Gamma]\ \dd p'\
+\ M\int_{-\infty}^{-q_{t+1}}\pproba(p'|p)\cdot[-g(t+1,p')-\Gamma]\
\dd p'$$
with $$\mc{H}=\int_{-\infty}^{+\infty}\pproba(p'|p)h(t+1,p')\dd p',$$
this gives us:
\begin{align*}
  V_t(\pi,p)&=\underset{|\pi'|\leq M}{\maximum}\ (\ g(t,p)\cdot\pi'-
  \Gamma|\pi-\pi'|+h(t, p)\ )\\
  &=\maximum\ [\ \underset{\pi\leq\pi'\leq M}{\maximum}\
  (g(t,p)-\Gamma)\pi'+\Gamma\pi\ ,\
  \underset{-M\leq\pi'\leq\pi}{\maximum}\
  (g(t,p)+\Gamma)\pi'-\Gamma\pi\ ]\ +\ h(t, p)
\end{align*}

\medskip

Using the definition of $g(t,p)$ above, we can prove the following~:
\begin{itemize}
\item $g(t,0)=0$ by symmetry
\item $g(t,p)$ is continuous and strictly increasing with $p$: indeed, if we
rewrite Equation~(\ref{equation_g}) as
$$ g(t, p)=p+\int_{-\infty}^{\infty}\pproba(p'|p)\ \gspecial(t+1,
  p')\textrm{d}p'$$
where $$\gspecial(t+1, p')=\begin{cases}
g(t+1, p')&\textrm{if }|p'|\leq q_{t+1}\\
\Gamma&\textrm{if }p'>q_{t+1}\\
-\Gamma&\textrm{if }p'<q_{t+1}
\end{cases}$$ then it suffices to see that $p\mapsto\gspecial(t+1,p)$
is continuous and increasing, as well as the cumulative distribution fuctions
$p\mapsto\pproba_{>q}(p)$ for $q\in\mathbb{R}$
%, because
%  this is true for $g(t+1,p)$, and the auto-correlation function of
%  the predictor is continuesly increasing
\item $\lim_{p\rightarrow+\infty}g(t,p)=+\infty$ because for $p>0$, we
  can show that ${\cal P}_{>q_{t+1}}(p)\geq{\cal P}_{<-q_{t+1}}(p)$
  and $\int_{-q_{t+1}}^{q_{t+1}}\pproba(p'|p)g(t+1, p')\geq 0$.
% thanks to symmetry and increasing auto-correlation.
\end{itemize}
Thus, there exists a unique $q_t\geq 0$ which satisfies
$$g(t, q_t)=\Gamma$$
and we finally have:
\begin{itemize}
\item if $p\geq q_t$, then $g(t,p)\geq\Gamma$ so:
$$V_t(\pi,p)=\maximum\ [\ \Delta M+\Gamma\pi\ ,\ (\Delta+\Gamma)\pi\ ]\ +\ h(t, p)$$
with $\Delta=g(t,p)-\Gamma$. But in order to have $\Delta M+\Gamma\pi
< (\Delta+\Gamma)\pi$ we need $\pi>M$ (or $\Delta=0$), so the maximum
is realised for $\pi'=M$, and $$V_t(\pi,p)=(g(t,p)-\Gamma)\cdot
M+\Gamma\pi+h(t,p)$$ Note that, technically, in the case $p=q_t$, then
$\Delta=0$ so that any $\pi'\geq\pi$ maximises $V_t(\pi,p)$: the
optimum is not unique anymore.  But for the sake of simplicity, we
impose the solution $\pi'=M$ in that particular case.
\item if $p\leq -q_t$, then $g(t,p)\leq-\Gamma$, and similarly we
  obtain a maximum for $\pi'=-M$,
  and $$V_t(\pi,p)=(-g(t,p)-\Gamma)\cdot M-\Gamma\pi+h(t,p)$$
\item if $|p|<q_t$, then $|g(t,p)|\leq\Gamma$, the maximum is realised
  for $\pi'=\pi$ and $$V_t(\pi,p)=g(t,p)\pi+h(t,p)$$
\end{itemize}\cqfd

\subsection{The self-consistency equation}\label{selfCons}

The solution given by Proposition~\ref{generalSolution} exhibits a
dependency in $t$.

Let us now consider the case where $T\rightarrow\infty$, and suppose
that the predictor is stationnary, i.e.: $\pproba(p_{t+1}=p'|p_t=p)$
is independent of $t$.  Then we obtain a \textit{telescopic} solution,
where the dependency in $t$ completely disappears, so that we only
have a one-variable function $g$ and a threshold $q^*$, satisfying the
following equations:

\medskip

\bec\framebox[12cm][c]{
\begin{minipage}{16cm}
\begin{gather}
  \label{selfConsEq}\qquad g(p)=p+\Gamma\cdot[{\cal P}_{>q^*}(p)-{\cal P}_{<-q^*}(p)]
  +\int_{-q^*}^{q^*}\pproba(p'|p)\ g(p')\ \textrm{d}p'\qquad\\
\label{equation_q}g(q^*)=\Gamma
\end{gather}
\vspace{-0.4cm}
\end{minipage}}
\eec

\medskip

Equation~(\ref{selfConsEq}) is a \textbf{self-consistent} functional
equation. The optimal solution to the system is then:
\begin{itemize}
\item if $p_t\geq \qstar$ then $\piSol_t=M$
\item if $p_t\leq -\qstar$ then $\piSol_t=-M$
\item if $|p_t|< \qstar$ then $\piSol_t=\piSol_{t-1}$.
\end{itemize}
Thus, we obtain a very simple trading system, always saturated at
$\pm\MaxPos$, with a threshold to decide at each step whether we
should revert the position or not.  This of course looks a lot like
the na\"\i ve solution from Section~\ref{naive}.  The only difference
lies in the value of the threshold $q^*$, defined by
Equations~(\ref{selfConsEq}) and~(\ref{equation_q}), instead of
$q_{\textrm{na\"\i ve}}=p_\infty^{-1}(\Gamma)$ for the na\"\i ve
solution. Intuitively, those equations take our future trading into
account, whereas the na\"ive solution does not.

If we look closely at Equation~(\ref{selfConsEq}), its interpretation
becomes transparent: $g(p)$ is equal
to $1/2M$ times the expected difference in total future profit between
the situation where $\pi=+M$ and the situation where $\pi=-M$. This
difference is made up of:
\begin{itemize}
\item the term $2 p M$ which represents the difference in immediate gain 
\item $2\Gamma M\cdot{\cal P}_{>q}(p)$ which represents the loss if
  the current position is $-M$ and in the next time step the predictor
  goes over the positive threshold $q^*$ (hence $\pi$ will go to $+M$)
\item $2\Gamma M\cdot{\cal P}_{<-q}(p)$ which represents the loss if
  the current position is $+M$ and in the next time step the predictor
  goes below the negative threshold $-q^*$ (hence $\pi$ will go to
  $-M$)
\item $\int_{-q}^{q}\pproba(p'|p)\ [2Mg(p')]\ \textrm{d}p'$ which is
  the expected difference in total future profit if, in the next step,
  the predictor remains between the two thresholds (leaving $\pi$
  unchanged).
\end{itemize}
Since the change of position between $-M$ and $+M$ costs $2 \Gamma M$,
it makes sense to compare $2M g(p)$ with it and only trade when $g(p)$
is greater than $\Gamma$. Hence, $g(p)$ can be seen as the ``gain per
traded lot''.

\medskip

Equation~(\ref{selfConsEq}) already allows us to say that $g(p)\geq
p$ for any $p\geq 0$ (and $g(p)\leq p$ for $p\leq 0$). As
$g(\qstar)=\Gamma$, this implies in particular that
$\qstar\leq\Gamma$. This property is actually rather intuitive:
indeed, if the immediate gain is higher than the trading cost, then
there is no reason not to trade the maximal possible amount.

\subsection{Reformulation as a path integral}\label{pathIntegral}

Although Equation~(\ref{selfConsEq}) is easy to interpret, it proves
very difficult to solve in concrete cases like the Ornstein-Uhlenbeck
case that we will consider in Section~\ref{ornstein}. In the present
section, we will extract an alternative equation for the optimal
threshold, which, although more sophisticated than the
self-consistency equation, will be easier to solve in practice.

\bigskip

Equation~(\ref{selfConsEq}) can be rewritten by expanding the function $g$:
\begin{align*}
  g(p)=&\ p+\int_{-\qstar}^{\qstar}
  p'\pproba(p'|p)\textrm{d}p'+\int_{-\qstar}^{\qstar}\int_{-\qstar}^{\qstar}
  p''\pproba(p''|p')\pproba(p'|p)\textrm{d}p'\textrm{d}p''+\dots\\
  &+\Gamma\cdot\left[\
    \int_{\qstar}^{+\infty}\pproba(p'|p)\textrm{d}p'
    +\int_{\qstar}^{+\infty}\int_{-\qstar}^{\qstar}
    \pproba(p''|p')\pproba(p'|p)\textrm{d}p'\textrm{d}p''+\dots\ \right]\\
  &-\Gamma\cdot\left[\
    \int_{-\infty}^{-\qstar}\pproba(p'|p)\textrm{d}p'+
    \int_{-\infty}^{-\qstar}\int_{-\qstar}^{\qstar}\pproba(p''|p')\pproba(p'|p)
    \textrm{d}p'\textrm{d}p''+\dots\ \right]
\end{align*}

Let us now consider that the predictor starts at $p_0=p$ at time $t$,
and follows the infinite path $(p_1, p_2, p_3, \dots)$ afterwards.
The above expansion tells us that this path will contribute to $g(p)$
as long as $-\qstar<p_i<\qstar$, and will stop contributing as soon as
$|p_i|\geq \qstar$ for some $i>0$.  Moreover, its contribution is
given by the sum $\sum_{i=0}^{n-1}p_i$, where $n>0$ is the first index
such that $|p_n|\geq \qstar$, and by $\pm\Gamma$, depending on whether
$p_n\geq \qstar$ or $p_n\leq -\qstar$.  We will only consider
predictors for which such an $n$ exists, which is true with
probability $1$ thanks to Condition~(\ref{unbound}) in
Section~\ref{description}: indeed, if $P_N$ is the probability for a
path of length $N$, starting at $p_0=p$, to satisfy $|p_i|<q$ for any
$1\leq i\leq N$, then $P_N\leq(1-\epsilon_q)^N$ with
$0<\epsilon_q\leq1$, so that $\lim_{N\rightarrow+\infty}P_N=0$.

If we now set $\pproba(p_0,\dots,p_n|p)=P(p_{t+i}=p_i,i\in[0,n]\ |\
p_t=p)$, this leads to the following equation:
\begin{align*}
  g(p)\ =&\ \sum_{n=0}^\infty\ \biggl[\
  \int_\qstar^{+\infty}\int_{-\qstar}^{\qstar}\dots\int_{-\qstar}^{\qstar}\
  \left(\sum_{i=0}^{n-1}\ p_i+\Gamma\right)\pproba(p_0,\dots,p_n|p)\
  \prod_{i=0}^n\textrm{d}p_i\\q
  &\qquad+\int_{-\infty}^{-\qstar}\int_{-\qstar}^{\qstar}\dots\int_{-\qstar}^{\qstar}\
  \left(\sum_{i=0}^{n-1}\ p_i-\Gamma\right)\pproba(p_0,\dots,p_n|p)\
  \prod_{i=0}^n\textrm{d}p_i\ \biggr]
\end{align*}
Using now the fact that $g(q^*)=\Gamma$, we get:
\begin{align*}
  \Gamma\ =&\ \sum_{n=0}^\infty\ \biggl[\
  \int_{q^*}^{+\infty}\int_{-q^*}^{q^*}\dots\int_{-q^*}^{q^*}
  \ \left(\sum_{i=0}^{n-1}\ p_i+\Gamma\right)\pproba(p_0,\dots,p_n|q^*)\ \prod_{i=0}^n\textrm{d}p_i\\
  &\qquad+\int_{-\infty}^{-q^*}\int_{-q*}^{q^*}\dots\int_{-q^*}^{q^*}\
  \left(\sum_{i=0}^{n-1}\ p_i-\Gamma\right)\pproba(p_0,\dots,p_n|q^*)\
  \prod_{i=0}^n\textrm{d}p_i\ \biggr]
\end{align*}

As we said above, there always exists, with probability $1$, an
integer $n$ such that $|p_n|\geq q$, so:
$$
1 = \sum_{n=0}^\infty\ \biggl[\ \int_{q^*}^{+\infty} +
\int_{-\infty}^{-q^*}\biggr] \int_{-q^*}^{q^*}\dots\int_{-q^*}^{q^*}
\pproba(p_0,\dots,p_n|q^*)\ \prod_{i=0}^n\textrm{d}p_i.
$$
which leads to:
\begin{equation*}
%\label{pathIntegralDiscrete}
  \sum_{n=0}^\infty\ \biggl[\ \int_{q^*}^{+\infty} +
  \int_{-\infty}^{-q^*}\biggr] \int_{-q^*}^{q^*}\dots\int_{-q^*}^{q^*}
  \left(\sum_{i=0}^{n-1}\ p_i-2\Gamma\cdot\indicFun_{\{p_n<-\qstar\}}\right)\pproba(p_0,\dots,p_n|q^*)
  \ \prod_{i=0}^n\textrm{d}p_i = 0
\end{equation*}
where $\indicFun$ is the indicator function.

This can be perhaps more gracefully expressed as a path integral: for
a finite path $\phi:[0,n]\rightarrow\mathbb{R}$, we note $T_\phi=n$,
$\phi_b=\phi(0)$, $\phi_e=\phi(n)$,
$\proba(\phi|p)=\pproba(p_{t+z}=\phi(z),\ z\in[0,n]\ |\ p_t=p)$ and
$\int_z\phi(z)\textrm{d}z=\sum_{i=0}^{n-1}\phi(i)$. The equation above
can then be symbolically expressed as:
\begin{equation}\label{pathIntegralFormula} 
  \boxed{\qquad\int\limits_{\substack{\phi_b=\qstar\\ -\qstar<\phi(z)<
        \qstar,\ z\in]0,T_\phi[}}^{|\phi_e|\geq
      \qstar}\ \left[\ \int_z\phi(z)\ \textrm{d}z -
      2\Gamma\cdot\mathbf{1}_{\{\phi_e\leq -\qstar\}}(\phi)\ \right]
    \ \pproba(\phi|\qstar)\ \pathInt\phi\ =\ 0\qquad}
\end{equation}

\begin{figure}[ht]
\bec\epsfig{file=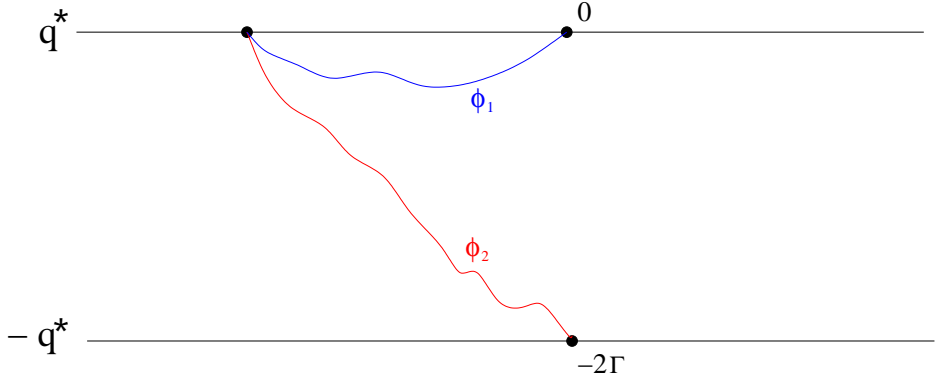, height=5cm}\eec
\caption{\label{pathIntegralFig}Path integral representation}
\end{figure}

Figure~\ref{pathIntegralFig} sums up this reformulation of the
problem: the value of $q^*$ is such that the ``penalty'' $2 \Gamma$
over all paths exiting through $-q^*$ is equal to the average gain
(given by the sum of the values of the predictor) over all paths
exiting either through $q^*$ or $-q^*$.

There is actually a direct interpretation for Equation
(\ref{pathIntegralFormula}), based on Figure~\ref{pathIntegralFig},
which is worth understanding. Imagine that we start with $p_t=p$ at
time $t$, and the position before our trading decision is
$\pi_{t-1}=-M$. Knowing how we will trade in the future (which depends
on the optimal threshold $\qstar$), we wonder whether it is worth
reverting the position right now by buying $\diffPos=2M$. Note that
the reasoning below does not depend on the actual value of $\diffPos$,
so one could as well consider buying just one share/contract.

% (the
%reasoning below would actually work for any value of $\diffPos$). 

Let us suppose that we do trade this quantity $\diffPos$, with a cost of
$\Gamma\cdot \diffPos$.  The next time we can possibly trade in the
future is the first time $T>t$ such that $|p_{T}|\geq\qstar$. If
$p_{T}\geq\qstar$ (path $\phi_1$ on Figure~\ref{pathIntegralFig}),
then we would have reverted our position anyway, so the cost of doing
it early can be considered as null, and our gain is given $\diffPos$
times the values taken by the predictor, hence
$\diffPos\int_z\phi(z)\textrm{d}z$.  If on the contrary we have
$p_{T}\leq-\qstar$ (path $\phi_2$ on Figure~\ref{pathIntegralFig}),
then we will revert again our position to $-M$ by paying
$\Gamma\cdot\diffPos$, hence a total cost of $2\Gamma\cdot\diffPos$,
whereas the gain will also be given by $\diffPos
\int_z\phi(z)\textrm{d}z$.

In the end, it is worth reverting the position to $\pi_t=M$ if, and
only if:
$$
\diffPos\ \cdot\int\limits_{\substack{\phi_b=p\\ -\qstar<\phi(z)< \qstar,\
    z\in]0,T_\phi[}}^{|\phi_e|\geq \qstar}\ \left[\ \int_z\phi(z)\
  \textrm{d}z -2\Gamma\cdot\mathbf{1}_{\{\phi_e\leq -\qstar\}}(\phi)\
\right]\ \pproba(\phi|\qstar)\ \pathInt\phi\ \geq\ 0$$ As the optimal
threshold determines exactly the limit between the trading zone and
the no-trading zone, we recover Equation~(\ref{pathIntegralFormula}).

\bigskip

Hence, the rather simple problem we introduced in the present article
has a very non-trivial solution, which is best described through the
above path-integral formulation.  Note that
Equation~(\ref{pathIntegralFormula}) is completely general provided
the assumptions of Section~\ref{description} are satisfied, it does
not rely on any specific statistics of the predictor.  In the next
section, we will explicitely solve this equation when the predictor is
Gaussian and follows a discrete Ornstein-Uhlenbeck evolution.

As a matter of notation, we will set:
\begin{align*}\ \pathIntPred(p)&=\int\limits_{\substack{\phi_b=p\\
      -\qstar<\phi(z)< \qstar,\ z\in]0,T_\phi[}}^{|\phi_e|\geq
    \qstar}\ \left[\ \int_z\phi(z)\ \textrm{d}z
    \ \right]\ \pproba(\phi|p)\ \pathInt\phi\\
  \probaCross(p)&=\int\limits_{\substack{\phi_b=p\\ -\qstar<\phi(z)<
      \qstar,\ z\in]0,T_\phi[}}^{\phi_e\leq -\qstar}\
  \pproba(\phi|p)\ \pathInt \phi,
\end{align*}
which can be interpreted, respectively, as the average contribution of
all paths before exiting the channel $[-q^*,q^*]$, and as the probability
for hitting the lower boundary $-q^*$ before the upper one $q^*$. In
terms of these quantities, Equation~(\ref{pathIntegralFormula}) now
writes:
\begin{equation}
\label{pathIntShort}
\pathIntPred(\qstar)=2\Gamma\cdot\probaCross(\qstar)
\end{equation}
In some cases, both sides of this Equation will tend to be infinitesimal,
so it is rather the ratio $\lim_{p \to \qstar} \pathIntPred(p)\ / \
\lim_{p \to \qstar} \probaCross(p)$ that we will ask to take the value
$2\Gamma$.

%\begin{equation}
%\label{pathIntShort}
%\frac{\lim_{p \to \qstar} \pathIntPred(p)}{\lim_{p \to \qstar} \probaCross(p)}= 2 \Gamma.
%\end{equation}
%Note that both the numerator and the denominator tend to zero when $p \to \qstar$, but their ratio stays finite in that limit.

\subsection{A note on non-linear predictability}\label{mostGeneralCase}

By looking at the proof of Proposition~\ref{generalSolution}, we note
that the linearity of the predictability is not a crucial
hypothesis. What we actually need is that
$\linearise_t(p_t)=E[r_t|p_t]$ satisfies the following properties, for
any $t$:
\begin{itemize}
\item $\linearise_t$ is continous and strictly increasing
\item $\lim_{p_t\rightarrow\infty}\linearise_t(p_t)>\Gamma$.
\end{itemize}

With these hypotheses it is possible to prove once again
Proposition~\ref{generalSolution}, except that
Equation~(\ref{equation_g}) becomes:
\begin{gather*}
  g(t, p)=\linearise_t(p)+\Gamma\cdot[{\cal P}_{>q_{t+1}}(p)-{\cal
    P}_{<-q_{t+1}}(p)]+\int_{-q_{t+1}}^{q_{t+1}}\pproba(p'|p)g(t+1,
  p')\textrm{d}p'
\end{gather*}
and the expression for $V_t(\pi, p)$ is similarly impacted.

If we want to consider the telescopic solution, then we need to have a
predictability independent of $t$, that is:
$\linearise_t(p)=\linearise(p)$.  This gives, for the self-consistency
equation:
\begin{gather*}
  g(p)=\linearise(p)+\Gamma\cdot[{\cal P}_{>q}(p)-
  {\cal P}_{<-q}(p)]+\int_{-q}^{q}\pproba(p'|p)\ g(p')\ \textrm{d}p'\\
  g(q)=\Gamma
\end{gather*}

This can again be solved using a path-integral formulation:
$$\int\limits_{\substack{\phi_b=\qstar\\ -\qstar<\phi(z)<
    \qstar,\ z\in]0,T_\phi[}}^{|\phi_e|\geq \qstar}\ \left[\
  \int_z\linearise(\phi(z))\ \textrm{d}z
  -2\Gamma\cdot\mathbf{1}_{\{\phi_e\leq -\qstar\}}(\phi)\ \right]\
\pproba(\phi|\qstar)\ \pathInt\phi\ =\ 0$$

\section{Application to an Ornstein-Uhlenbeck predictor}\label{ornstein}

\subsection{Definition}

We will now focus on the case of a predictor following a discrete
Ornstein-Uhlenbeck dynamics:
\begin{gather}\label{ornsteinUhlenbeck}
p_{t+1}-p_t=-\epsilon\cdot p_t+\beta\cdot \xi_t
\end{gather}
where $(\xi_t)_{t\in\mathbb{R}}$ is a set of independent $\mc{N}(0,
1)$ Gaussian random variables.

\bigskip

One classical example of such a
predictor is an exponential moving average of price returns:
$$
p^{EMA}_t=K\sum_{t'<t}\rho^{t'-t-1} r_t
$$
If we suppose, as is usual, that the returns $r_t$ are $\mc{N}(0,
\sigma_r)$ random variables, then this gives an Ornstein-Uhlenbeck
predictor with $\epsilon=1-\rho$ and $\beta=K\sigma_r$. Note however
that the $r_t$ must have some small correlations in order to be
predictable!  Therefore, in this case, the discussion in
terms of an Ornstein-Uhlenbeck process is only consistent in
the limit of small predictability, i.e. $K \ll 1$.

% whole discussion in terms of an Ornstein-Uhlenbeck
%process is therefore only consistent in the limit of small
%predictability, i.e. $K \ll 1$.

\subsection{Properties \& Orders of magnitude}

Let us consider Equation~(\ref{ornsteinUhlenbeck}) with the hypothesis
that $\epsilon \ll 1$. Then we have $p_{t+1} \approx e^{-\epsilon} p_t
+ \beta\cdot \xi_t,$ so that
$$E[p_{t+n}|p_t]\approx e^{-\epsilon n}p_t.$$
So, $\tau=\epsilon^{-1}$ is the \textbf{auto-correlation time} of
the predictor $p_t$. The standard deviation of the predictor,
\ie\ its \textbf{average predictability}, $\sigma_p=\sqrt{E[p_t^2]}$,
is given by $\beta/\sqrt{2\epsilon}$ (in the limit $\epsilon \ll 1$).

Hence:
\begin{itemize}
\item the smaller $\epsilon$ is, the longer the predictor
  takes to express itself
\item the higher $\beta$ is, the better the signal is (on
  average).
\end{itemize}
The integrated predictability is given by
\begin{align*}
p_\infty(p) &= \sum_{n=0}^\infty E[p_{t+n}|p_t] \approx \sum_{n=0}^\infty e^{-\epsilon n}p_t\\
&\approx p /\epsilon.
\end{align*}
This implies that the na\"ive threshold value is given by
$q_{\textrm{na\"ive}}=\Gamma\epsilon$, while the integrated average
predictability is:
$$\sigma_\infty=\frac{\beta}{\sqrt{2 \epsilon^3}}$$

In practice, if a real price predictor is to be both useful and
realistic, it should beat the trading costs when the predictor value
is a few times its standard deviations. This allows to obtain a system
which trades regularly, but not too often, compared to its
auto-correlation time.  For our Ornstein-Uhlenbeck predictor, this
implies that when $p_t\propto \beta/\sqrt{2\epsilon}$, one should also
have $p_t \epsilon^{-1}\propto\Gamma$.  Therefore, the interesting
regime for practical applications is:
$$
\beta\propto \Gamma\epsilon^{3/2}.
$$

In what follows, we will study the problem by distinguishing between two cases:
\begin{itemize}
\item If $\beta\gg\Gamma$, the predictor can easily beat its
  transaction costs at every step. This situation (which is not very
  realistic) requires us to keep a \textbf{discrete time} approach of
  the problem.
\item If $\beta\ll\Gamma$, the predictor needs in general a large
  number of steps to beat the costs. This will lead us to a
  \textbf{continuous} formulation (and resolution) of the problem.
\end{itemize}

\subsection{Discrete case:  $\beta\gg\Gamma$}\label{discreteCase}

We already explained in Section~\ref{selfCons} that we always have $q^*\leq\Gamma$.
Consequently, whenever
$\beta\gg\Gamma$, we also have $\beta\gg q^*$. This means that,
starting at $p=q^*$, one will typically jump beyond $q^*$ or $-q^*$ in
just one step. Thus:
\begin{align*}
  \pathIntPred(q^*)&=q^*\\
  \probaCross(q^*)&=\int_{x^*}^{+\infty}\frac{e^{-x^2/2}}{\sqrt{2\pi}}
  \textrm{d}x\qquad\textrm{with}\qquad
  x^*=\frac{(2-\epsilon)q^*}{\beta}
\end{align*}
Since $\beta\gg q^*$, one has $x^*\ll 1$, and thus
$\probaCross(q^*)\approx 1/2$. Equation~(\ref{pathIntShort}) finally
gives:
\begin{equation}
q^*=\Gamma
\end{equation}
Hence, if the volatility of each predictor change is very large
compared to the trading costs, then one needs to be as selective as
possible.

\subsection{Continuous case: $\beta\ll\Gamma$}\label{continuous}

First, let us show why the condition $\beta\ll\Gamma$ leads us to
express the problem in a continuous form. Under that condition, we
cannot have an optimal threshold $q^*$ of the same order as magnitude
as $\beta$ itself.  Indeed, if this was the case, any time the
predictor has the value $q^*$, it would have a significant probability
to go below $-q^*$ at the next step since the predictor changes by an
amount $\propto \beta$ at each time step.  The optimal strategy would
then require to resell everything at cost $2\Gamma$, whereas the
immediate gain would only be of the order of magnitude of
$\beta$. Therefore, Equation~(\ref{pathIntShort}) could not be
satisfied.

Now, knowing that $q^*\gg\beta$, we need to evaluate
$\probaCross(q^*)$ and $\pathIntPred(q^*)$. But for the predictor to
go from $q^*\gg\beta$ to $-q^*\ll-\beta$ requires many
steps. Therefore, one is effectively is the continuum limit, where the
variation of the predictor at each time step is infinitesimal compared
to $q^*$. We can then approximate the dynamics of the predictor
by a drift-diffusion process:
\begin{equation}
\label{driftDiffusion}\textrm{d}p=-\epsilon p\ \textrm{d}t + \beta\ \textrm{d}X_t
\end{equation}
where $(X_t)_t$ is a Wiener process.

In such a continuous setting, the quantities $\pathIntPred(q^*)$ and
$\probaCross(q^*)$ are actually ill-defined because the diffusion
process starts on an absorbing boundary.  This is a classical problem,
which is handled by starting infinitesimally close to $q^*$. Therefore
we consider $\pathIntPred(p)$ and $\probaCross(p)$ for $p < q^*$.  It
is easy to show that these two functions obey two \textbf{Kolmogorov
  backward equations}, that read:
\begin{equation}\label{equaDiffL}
  \boxed{\qquad\frac{1}{2}\beta^2\ \frac{\partial^2\pathIntPred}{\partial p^2}\ -
    \ \epsilon p\ \frac{\partial\pathIntPred}{\partial p}\ =\ -p\ ; \qquad
    \frac{1}{2}\beta^2\ \frac{\partial^2\probaCross}{\partial p^2}\ -
    \ \epsilon p\ \frac{\partial\probaCross}{\partial p}\ =\ 0\qquad}
\end{equation}
with boundary conditions: $\pathIntPred(\pm q^*)=0$ and
$\probaCross(q^*)=0$, $\probaCross(-q^*)=1$.

\subsubsection{Solution}

Solving Equations~(\ref{equaDiffL}) with their boundary conditions
leads to:
\begin{align*}
  \pathIntPred(p)&=\frac{1}{\epsilon}\left(p-\frac{q}{I}\int_0^pe^{av^2}\textrm{d}v\right)\\
  \probaCross(p)&=\frac{1}{2}\left(1-\frac{1}{I}\int_0^pe^{av^2}\textrm{d}v\right)\end{align*}
with
$$
I=\int_0^{q^*}e^{av^2}\textrm{d}v\qquad\textrm{and}\qquad
a=\frac{\epsilon}{\beta^2}.
$$

Setting now $p=q^*-u$ with $u \to 0$, one finds that
Equation~(\ref{pathIntShort}) becomes, to first order in $u$:
$$
-\frac{u}{\epsilon}+\frac{uq^*}{\epsilon}\cdot \frac{e^{aq^{*2}}}{I}\
\approx \ \Gamma u\ \frac{e^{aq^{*2}}}{I}.
$$
As expected, $u$ disappears from the equation, to give the following
solution for the threshold $q^*$:
\begin{equation}
  \label{finalEq}\boxed{\qquad q^*=\frac{\beta}{\sqrt{\epsilon}}
    \ F^{-1}\left(\frac{\Gamma\epsilon^{3/2}}{\beta}\right)\qquad\textrm{where}
    \qquad F(x)=x-e^{-x^2}\int_0^xe^{v^2}\textrm{d}v\qquad}
\end{equation}
Note that when $\epsilon\ll 1$, this equation can be expressed
entirely in terms of the integrated predictability:
$$
p_\infty(q^*)=\Gamma\cdot
H\left(\frac{\sigma_{\infty}\sqrt{2}}{\Gamma}\right)\qquad\textrm{where}\qquad
H(x)=x\ F^{-1}\left(\frac{1}{x}\right).
$$
This means that we can find the optimal threshold for a predictor by
studying only its total predictive power (if we suppose of course that
it satisfies all the required properties).

\subsubsection{Limits}

One can now study the limits of Equation~(\ref{finalEq}) for large and
small values of the only remaining adimensional parameter
$\eta={\Gamma\epsilon^{3/2}}/{\beta}$. Interestingly, $\eta \sim 1$ is
the regime mentioned above where predictability beats costs whenever the
predictor's value is of the order of its rms.

The limiting behaviours of the function $F(x)$ are as follows:
\begin{itemize}
\item if $x \gg 1$, then $\int_0^xe^{v^2}\textrm{d}v \ll e^{x^2}$, so
  $F(x)\ \approx\ x$
\item if $x \ll 1$, then $F(x)\ \simeq\
  x-(1-x^2)\int_0^x(1-v^2)\textrm{d}v\ \approx \ \frac{2x^3}{3}$.
\end{itemize}
Therefore when $\eta \gg 1$, the threshold is simply given by $q^* =
\Gamma \epsilon$. This result is rather intuitive: if $\beta$ is very
small then the predictability of the predictor is weak, compared to
the trading cost. Hence, it makes sense to try to catch any profitable
opportunity, without taking future trading into account.  That is why
we recover the na\"\i ve solution of Section~(\ref{naive}).

If on the other hand $\beta \gg \Gamma\epsilon^{3/2}$ then $\eta \ll
1$, and $F^{-1}(\eta)\ \approx\ \sqrt[3]{\frac{3}{2}\cdot \eta}$,
which finally gives
\begin{equation}
q^*=\sqrt[3]{\frac{3}{2}\cdot\Gamma\beta^2}
\end{equation}
This is the result that would obtain with a predictor following a
Brownian motion. Indeed, if $\beta$ is large enough, the
mean-reverting effect $\epsilon$ is not relevant, and the optimal
threshold must consequently be independent of $\epsilon$.

\subsection{Shape of the global solution}

\begin{figure}[ht]
\bec\epsfig{file=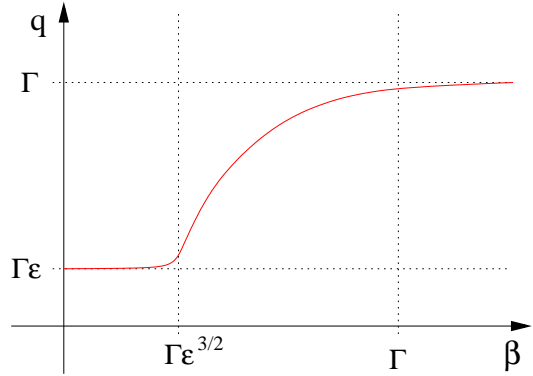, height=5cm}\eec
\caption{\label{figSolution}Optimal thershold as a function of
  $\beta$}
\end{figure}

We summarize the various regimes in Figure~\ref{figSolution}, where we
plot the optimal threshold $q^*$ as a function of $\beta$, for
$\epsilon$ and $\Gamma$ fixed.  One can see the three main
caracteristics of this solution:
\begin{itemize}
\item a constant threshold $q^*=\Gamma\epsilon$ for small values of $\beta \ll \Gamma\epsilon^{3/2}$
\item a sublinear behaviour $q^* \propto \Gamma^{1/3} \beta^{2/3}$ in
  the intermediate regime between $\beta \sim \Gamma\epsilon^{3/2}$
  and $\beta \sim\Gamma$
\item a constant threshold $q^*=\Gamma$ for large values of $\beta \gg \Gamma$.
\end{itemize}

Note that these different results match seamlessly at the boundaries
between the regimes. Indeed,
$q^*=\sqrt[3]{\frac{3}{2}\cdot\Gamma\beta^2}$ becomes of the order of
$\Gamma\epsilon$ when $\beta\sim \Gamma\epsilon^{3/2}$, and becomes of
of $\Gamma$ when $\beta \sim \Gamma$ is of the order of $\Gamma$. One
can also compute systematic corrections to $q^* = \Gamma$ as an
expansion in $\Gamma/\beta$:
$$
q^*\ \approx\ \Gamma\ -\ (1 - \epsilon) \sqrt{\frac{2}{\pi}}\ \cdot\ \frac{\Gamma^2}{\beta}\ +\ \dots \qquad \beta \gg \Gamma.
$$

\subsection{The case of a white noise predictor}

To conclude, let us consider the special case of
Equation~(\ref{ornsteinUhlenbeck}) where $\epsilon = 1$. In that case,
the predictor is a white noise in time: $E[\ p_tp_{t+1}\ ]=0$.  Since
we assume a perfect $p \mapsto -p$ symmetry, the self-consistency equation
becomes simply $g(p)=p$, which trivially implies that $q^*=\Gamma$ in
that case. This is also consistent with our explicit solutions above:
when $\epsilon=1$, the intermediate regime disappears and one indeed
finds $q^*=\Gamma\epsilon=\Gamma$.  This threshold is what we expect
from such a system: without any auto-correlation, the best strategy is
to trade as soon as the instantaneous predictability is above the
trading cost.  Note that in this case $p_\infty=p_t$, so this
threshold also coincides with the na\"\i ve solution.

\section{Numerical results}

To check the robustness of our analytical results, we ran some
simulations to determine the optimal threshold numerically, to be
compared with the theoretical value we obtained in
Section~\ref{ornstein}.

In what follows we set $\epsilon=0.001$. We can set $\Gamma=1$ without
loss of generality, and the only remaining variable is $\beta$.

\begin{figure}[ht]
\bec\epsfig{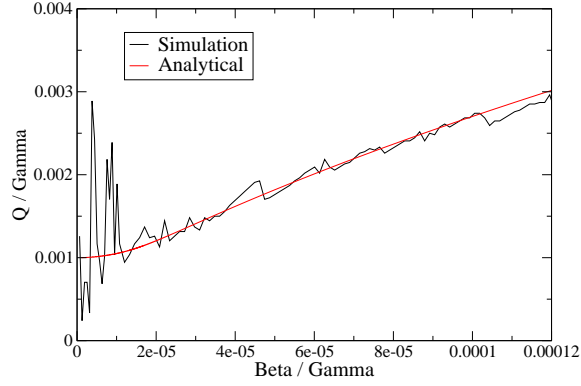}\eec
\caption{\label{simuOptTh1}Optimal threshold for $0< \beta< 1.2\cdot10^{-4}$}
\end{figure}

\begin{figure}[ht]
\bigskip
\bigskip
\bec\epsfig{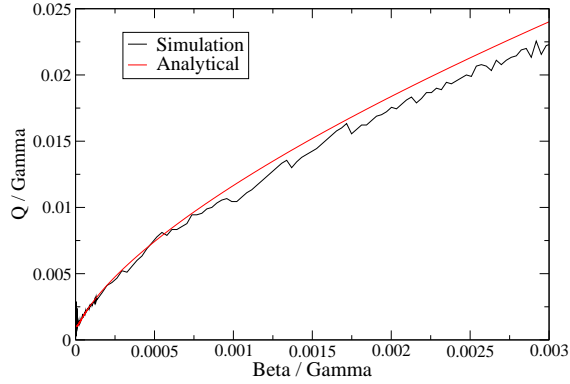}\eec
\caption{\label{simuOptTh2}Optimal threshold for $0< \beta< 0.003$}
\end{figure}

The algorithm to find the optimal threshold for a given value of
$\beta$ runs as follows:
\begin{enumerate}[1)]
\item We choose a set of threshold values $q_1,\dots,q_n$ uniformely
  distributed over a reasonably large range (which contains the
  theoretical optimal threshold).
\item We generate a long random path $(p_t)_{t\in[0,T]}$ for the
  predictor, following the law given by
  Equation~(\ref{ornsteinUhlenbeck}).
\item For each value of the threshold, we simulate the behaviour of
  the corresponding strategy, with a \MaxPos\ of 1.
\item To obtain the P\&L of each system, we calculate at each time
  step $t$ the gain given by\footnote{Indeed, as we only consider
    expected values, there is no need to generate a random variable
    for the price return $r_t=\Price_{t+1}-\Price_{t}$ as a function
    of the predictor $p_t$~: one can directly consider the mean of
    this variable, which is exactly $p_t$.}  $p_t\cdot\pi_t$ (where
  $\pi_t$ is the position of the system) and the cost given by
  $\Gamma|\delta{\pi_t}|$ if there is a trade $\delta{\pi_t}$.
\item We select the threshold $q_{j}$ with the maximal total P\&L.
\item We choose new values $q'_1,\dots,q'_n$ for the threshold,
  distributed around $q_{j}$, and we restart the algorithm with these
  values.
\end{enumerate}
This loop is repeated several times, in order to get a sufficiently
precise approximation of the optimal threshold.  The result of this
process is shown in Figures~\ref{simuOptTh1} and~\ref{simuOptTh2}. The
analytical solution in the continuous case, given by
Equation~(\ref{finalEq}), is easy to compute (the function
$D(x)=e^{-x^2}\int_0^xe^{v^2}\textrm{d}v$ is a classic, called the
\textbf{Dawson function}).

By comparing the two curves, one can check that the analytical
solution is indeed a good fit of the simulation results. However,
there are two interesting details to note:
\begin{itemize}
\item when the value of $\beta$ becomes very small
  (Figure~\ref{simuOptTh1}), the solution of the simulation becomes
  very noisy;
\item there is a discrepancy between the analytical and simulated
  solutions when $\beta$ increases (Figure~\ref{simuOptTh2}) and,
  rather surprisingly, this happens in a regime where the inequality
  $\beta\ll\Gamma$ still seems to hold.
\end{itemize}

The reason for the first effect is obvious: when $\beta$ becomes very
small, then the predictor hardly ever beats its optimal threshold
$\Gamma\epsilon$ during the course of the simulation, which implies
that the real optimal threshold becomes very difficult to find: we
need to use a huge number of steps in the simulation to find the
proper solution.

The answer to the second remark is more subtle: if we go back to
Section~\ref{continuous}, we see that it is not really the hypothesis
$\beta\ll\Gamma$ which implies that we are in a continuous setting,
but rather the inequality $\beta\ll q^*$. Of course this second
inequality was derived from the first one, but we did not specify at
that time the orders of magnitude involved in each inequality. In
fact, if we consider for the threshold $q^*$ the limit value
$\sqrt[3]{\frac{3}{2}\cdot\Gamma\beta^2}$, one sees that
$\frac{q^*}{\beta}=\kappa\ \sqrt[3]{\frac{\beta}{\Gamma}}$ with
$\kappa\approx 1.15$.  This means that our so-called continuum
hypothesis breaks down much sooner than a na\"\i ve comparison between
$\beta$ and $\Gamma$ would tell. In Figure~\ref{simuOptTh2}, if we
look at the point for which $\beta/\Gamma \approx 0.002$, one actually
has $q^*/\Gamma \approx 0.144$; so the continuum hypothesis is already
unwarranted for this value, and the solutions do not strictly
coincide, even though the ratio between $\beta$ and $\Gamma$ is still
extremely small.

\section{Extension to other risk constraints}

\subsection{The ``band'' system}

The problem we have presented and solved in the previous sections was
rather specific, as it required a risk control based on the only
constraint $|\pi|\leq M$. An alternative, more classical way to handle
risk, is to consider a quadratic penalty $R(\pi_t)= \pi_t^2$ that
represents risk aversion. The ``utility'' to be maximized is at each
step $t$ given by $g_t=r_t\pi_t-\lambda \pi_t^2$. Note that the
penalty term could read, more generally, $R(\pi_t) \propto
|\pi_t/M|^z$. The case $z=2$ is the above quadratic penalty, whereas
the constraint on the maximum position formally corresponds to the
limit $z \to \infty$.

The quadratic problem was considered in~\cite{meanReversion,
  martinMulti}, where it was showed that the optimal strategy in this
setting is a \textbf{band}, also called a DT-NT-DT (Direct Trading -
No Trading - Direct Trading) system. This system is defined as
follows:
\begin{itemize}
  % \item any position $\pi$ defines an ``image predictor''
  %   $p=2\lambda\cdot\pi$ and a band of size $2\qstar$ around that
  %   value (and reciprocally, any predictor $\pi$ defines an ``ideal
  %   position'' $\pi=p/ 2\lambda$)
  % \item at each time step, if the image of the current position is
  %   inside the band then there is no trade; but if it is outside of
  %   the band then the trade should bring the image of the position
  %   to the closest border of the band (i.e. the new position should
  %   be either $\pi + \qstar/ 2\lambda$ or $\pi - \qstar/ 2\lambda$).
%\item any predictor $p$ defines an ``ideal position'' $\pi=2p/\lambda$,
%with a band of size $4\qstar/\lambda$ around that value
% \item at each time step, if the current position is inside the band
%   then there is no trade; but if it is outside of the band then the
%   trade should bring the position to closest border of the band
%   (i.e. either $\pi + 2\qstar/\lambda$ or $\pi - 2\qstar/\lambda$).
\item there is a one-to-one correspondence between the values of the
  predictor and the position (in the quadratic case, this is simply
  given by: $\pi=p/ 2\lambda$)
\item at each time step, one defines a band of size $2\qstar$ around
  the value of the predictor
\item if the image of the current position is inside the band then
  there is no trade; but if it is outside of the band then the trade
  should bring the image of the position to closest border of the
  band.
% (i.e. either $\pi + 2\qstar/\lambda$ or $\pi - 2\qstar/\lambda$).
\end{itemize}

Note that we did not explicit here the position of the predictor $p_t$
inside the band. It is actually proven in~\cite{meanReversion} that it
does not have to be at the middle of the band, but it does when the
trading cost becomes small. The band will be called \textbf{symmetric}
in this case. This symmetric band is a property that will be necessary
for our argument below to work.

\bigskip

This band system looks at first sight rather remote from the system we
studied in the previous sections. Still it was shown in
\cite{meanReversion} that if the asset price follows a mean-reverting
dynamics (which corresponds exactly to the case of a continuous
Ornstein-Uhlenbeck predictor) then the optimal half-size of the band
is given, with the notations of the present paper, by:
$$\qstar=\sqrt[3]{\frac{3}{2}\cdot\Gamma\beta^2}$$
when the trading cost $\Gamma$ is small. This is exactly the value of
our threshold in the continuous setting, when costs are small!

Our goal in the next section is to explain that this in no
coincidence.  This will allow us not only to recover the results
of~\cite{meanReversion}, but also to extend them by giving the optimal
solution under the condition that the band is symmetric.

\subsection{Optimal size of the band}

In what follows, we suppose that we already know that the optimal
trading policy is a band system as described above, with the predictor
being at the center of the band, and we will find the value of the
half-band for this optimal system, under some condition on the current value
of the predictor.

Consider that we start with $p_t=\pRef$, and the position
before the next trading decision is at the lower border of the band:
$\pi_{t-1}=(\pRef-\qstar)/2\lambda$.  Knowing that the future trading
style is a band of optimal size $2\qstar$, we wonder if it is worth
buying an infinitesimal quantity $\delta\pi$ at $t$.

\begin{figure}[ht]
\bec\epsfig{file=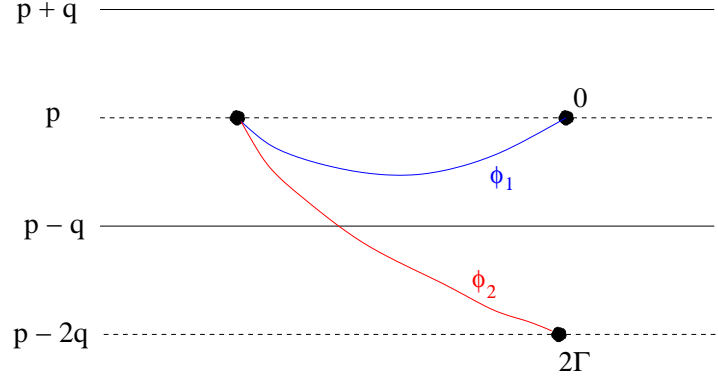, height=5cm}\eec
\caption{\label{figPbBand}Paths in the band system. The band is the
  region between $p-q$ and $p+q$, the paths $\phi_1$ and $\phi_2$ are
  the predictors trajectories defining the two possible ways to get
  out of the band if we start at the lower border.}
\end{figure}

Let us suppose that we buy this quantity. The new risk penalty term
will then be
\begin{align*}
R(\pi_t)&=\lambda(\pi_{t-1}+\delta\pi)^2\\
&=R(\pi_{t-1})+(\pRef-\qstar)\cdot\delta\pi+O(\delta\pi^2)
\end{align*}

Now let us follow the scenarii given by Figure~\ref{figPbBand}: by
definition of the band system, we will not trade until the predictor's
value becomes larger than $\pRef$ or smaller than $\pRef-2\qstar$. In
the former case, then the trading cost is zero because we would have
bought $\delta\pi$ anyway, whereas in the latter case we will have to
sell it back, which gives a total cost of $2\Gamma\cdot\delta\pi$.

In both cases, if we note $\phi$ the path between $t$ and the first
$T$ such that $p_T\geq\pRef$ or $p_T\leq\pRef-2\qstar$, then the gain
of trading $\delta\pi$ is given by
$$\mathcal{G}_{\delta\pi}=\int_z(\phi(z)\cdot\delta\pi-\delta R)\ \textrm{d}z$$
where $\delta R=R(\pi_t)-R(\pi_{t-1})\simeq(\pRef-\qstar)\cdot\delta\pi$.

In the end, we have a positive gain by trading $\delta\pi$ if, and
only if:
$$
\delta\pi\ \cdot\int\limits_{\substack{\phi_b=\pRef\\
    \pRef-2\qstar<\phi(z)< \pRef,\
    z\in]0,T_\phi[}}^{|\phi_e-\pRef+\qstar|\geq \qstar}\ \left[\
  \int_z(\phi(z)-\pRef+\qstar)\ \textrm{d}z
  -2\Gamma\cdot\mathbf{1}_{\{\phi_e\leq \pRef-2\qstar\}}(\phi)\
\right]\ \pproba(\phi|\pRef)\ \pathInt\phi\ \geq\ 0$$

We can then do the change of variable $\psi(z)=\phi(z)-\pRef+\qstar$,
but in order to have $\pproba(\psi|\qstar)=\pproba(\phi|\pRef)$ we
need to be in a case when the drift of the predictor can be neglected, i.e. when $\epsilon$ 
is small enough: $\qstar\ll\beta/\sqrt{\epsilon}$.
%By a change of variable $\psi(z)=\phi(z)-\pRef+\qstar$,
This gives (as $\delta\pi>0$):
$$
\int\limits_{\substack{\psi_b=\qstar\\ -\qstar<\psi(z)<
      \qstar,\ z\in]0,T_\psi[}}^{|\psi_e|\geq
    \qstar}\ \left[\ \int_z\psi(z)\ \textrm{d}z
  -2\Gamma\cdot\mathbf{1}_{\{\psi_e\leq
    -\qstar\}}(\psi)\ \right]\ \pproba(\psi|\qstar)\
\pathInt\psi\ \geq\ 0
$$

The limit of the no-trading zone is the point where the cost of
trading equilibrates precisely the gain, so the above inequality
becomes an equality, and we recover exactly
Equation~(\ref{pathIntegralFormula})!  This means that the same
equation defines the optimal threshold in the quadratic risk-control
case with small trading costs and in the ``MaxPos'' setting.

\subsection{Consequences}

As a consequence of our above result, we recover the optimal half-band
size $\qstar=\sqrt[3]{\frac{3}{2}\cdot\Gamma\beta^2}$ for small values
of $\Gamma$, and in particular the two-third dependency on the costs
explained in~\cite{delta23}. This also gives us the optimal value for
any $\Gamma$, provided that the band is symmetric and
$\qstar \ll\beta/\sqrt{\epsilon}$: this value will be given by
Equation~\ref{finalEq}.  This can be seen as a ``variational
solution'', where we get the optimal solution in a close-to-optimal
subspace of the space of possible trading systems. But in reality, as 
shown in ~\cite{meanReversion}, there is a small shift 
of the center of the band, which is however of higher order in $\Gamma^{1/3}$.

Moreover, our result is more general than the case of an
Ornstein-Uhlenbeck predictor: Equation~(\ref{pathIntegralFormula})
works for any price predictor, whatever its dynamics, provided it
satisfies the hypotheses of Section~\ref{description}.  And the
technique based on Kolmogorov backward equations, that we presented in
Section~\ref{continuous}, can easily be extended to these various
dynamics.

Finally, it can be proven that for any reasonable risk-function
$R(\pi)$ -- for example $R(\pi) = |\pi|^z$ for any $z > 0$, if the
trading system is a symmetric band then its half-size is again given
by Equation~(\ref{pathIntegralFormula}).  The limit $z \to \infty$ is
singular in the sense that one loses the one-to-one correspondance
between the predictor and the size of the trade, but formally our
result holds for arbitrary $z$, and therefore applies to the
``MaxPos'' system considered in the previous sections.

\section{Conclusion}

We have considered and solved exactly the problem of the optimal
trading strategy when one wants to follow a completely general
Markovian predictor of the future returns of a single asset, in the
presence of linear costs, and with a strict cap on the allowed
position in the market. Using Bellman's backward recursion method, we
have shown that the optimal strategy is to switch between the maximum
allowed long position and the maximum allowed short position, whenever
the predictor exceeds a {\it threshold value}, for which
we establish an exact, non-trivial, equation. This equation can be
solved explicitely in the case of a discrete Ornstein-Uhlenbeck
predictor. We discussed in detail the dependence of this threshold
value on the transaction costs.

We also showed an unexpected relation between our problem and the
problem where risk is handled dynamically, with an arbitrary risk
penalty.  The connection relies on the presence of a no-trading zone for
the two problems, which allows a use of powerful Bellman techniques
to calculate the optimal parameter.

% This led us not only to recover the known one-third power dependency
% of the size of the band in the cost parameter, but also to give an
% explicit general solution to the problem for arbitrary Markovian
% predictors and arbitrary risk penalty functions.

There are various interesting extensions of our results that one can
think of. One could consider the case of a predictor with jumps, and
see how this affects the threshold value. One could consider an
asymmetric risk constraint, where the maximum long and short positions
are different (this could be relevant for option trading).  But the
most relevant extension would be to consider the case, important in
practice, where the costs have both a linear component (coming from
fees, bid-ask spread, etc.) and a quadratic component that would model
impact. One may hope that an exact solution is still available in some regime, 
at least in the case of a single asset, or when the risk
constraints do not couple different assets.

\section*{Acknowledgements} We thank Nicolas Bercot and Julianus
Kockelkoren for fruitful discussions, Emeric Henry for reading the 
manuscript and Richard Martin for
his precious comments and references.

\bibliographystyle{alpha}
\bibliography{ll}

\end{document}